\documentclass[useAMS,usenatbib]{mn2e}
\usepackage{graphicx}

\def\aj{AJ}%
\def\apj{ApJ}%
\def\aap{A\&A}%
\def\mnras{MNRAS}%

\title[Oxygen, $\alpha$-element and iron abundance in the inner part of the Galactic disc. II]
{Oxygen, $\alpha$-element and iron abundance distributions in the inner part of the Galactic 
thin disc. II \thanks{Based on observations obtained at the Canada-France-Hawaii Telescope
(CFHT), which is operated by the National Research Council of Canada, the
Institut National des Sciences de l'Univers of the Centre National de la
Recherche Scientifique of France, and the University of Hawaii.}}

\author[S.M. Andrievsky et al.]
{S.M. Andrievsky$^{1,2}$,
R.P. Martin$^{3}$,
V.V. Kovtyukh$^{1}$,  
S.A. Korotin$^{1}$,
J.R.D. L\'epine$^{4}$\\ 
$^{1}$Department of Astronomy and Astronomical Observatory, Odessa
National University\\ and Isaac Newton Institute of Chile, 
Odessa Branch, Shevchenko Park, 65014 Odessa, Ukraine\\
email:andriecskii@ukr.net\\
$^{2}$GEPI, Observatoire de Paris-Meudon, CNRS, Universite Paris Diderot, 92125
Meudon Cedex, France\\
$^{3}$Department of Physics and Astronomy, University of Hawai'i at 
Hilo, Hilo, HI, 96720, USA\\
$^{4}$Universidade de S\~ao Paulo, IAG/USP - Departamento de Astronomia, 
Rua do Mat\~ao 1226 - Cidade Universitaria, 05508-900 - Sao Paulo, SP\\
}

\begin{document}

\date{Received ; accepted }

\maketitle

\begin{abstract}

We have derived the abundances of 36 chemical elements in one Cepheid star, ASAS 181024--2049.6, located 
R$_{\rm G}= 2.53$ kpc from the Galactic center. This star falls within a region of the inner thin disc poorly sampled in 
Cepheids. Our spectral analysis shows that iron, magnesium, silicon, calcium and titanium LTE abundances 
in that star support the presence of a plateau-like abundance distribution in the thin disc within 5 kpc of the Galactic 
center, as previously suggested by \cite{Maret15}. If confirmed, the flattening of the abundance gradient within that region 
could be the result of a decrease in the star formation rate due to dynamic effects, possibly from the central Galactic bar.
\end{abstract}


%

\section{Introduction}

In our recent paper on oxygen and $\alpha$-element abundances derived from the spectra of Cepheids 
situated in the vicinity of the Galactic center  \citep{Maret15}, we showed that the abundance distribution 
across the Galaxy is complex. Globally, chemical abundances increase from the outer part of the Galactic disc towards the 
Galactic center. However, the distribution reaches an apparent maximum at about R$_{\rm G}$ $\approx$ 5 kpc, forming a kind 
of plateau-like structure, and with a possible gradual decrease of abundances toward the inner central zone. Nevertheless, 
this flattening in the abundance distribution in the inner Galactic disc has not been firmly demonstrated yet since our study 
was incomplete to some extent: in that earlier paper, no Cepheids were studied between the Galactic center and a distance 
of R$_{\rm G}$ $\approx$ 3 kpc. We also relied on other spectral data from the literature to assess the abundances 
at the very center of the Galaxy. 

It is difficult to find Cepheid stars in the inner disc for conducting our spectroscopic 
analysis. But, after a careful inspection of the literature data on new Cepheid candidates situated toward 
the Galactic center, we have selected four possible stars listed in Table \ref{Obs}. \\

\begin{table}
\small
\caption{List of observed stars}
\label{Obs}
\begin{tabular}{crrllr}
\hline
  ASAS Star              &  P, d & $V$  &    RA    &      DEC   &  R$_{\rm G}$ [kpc] \\
\hline
175621-2738.0  &    16.33& 11.69  & 17 56 21.4& -27 37 55  & 1.93 \\
181024-2049.6  &    24.43& 13.52  & 18 10 23.7& -20 49 28 & 2.53 \\
183406-1519.1  &     9.22&  9.24   & 18 34 05.4& -15 19 10  & 2.70 \\
193435-1921.7  &     2.12& 13.02  & 19 34 34.6& -19 21 40  & 2.93 \\
\hline
\end{tabular}
\end{table}


\section{Observations and data reduction}

We observed our stars with the 3.6-m Canada--France--Hawaii Telescope (CFHT), 
on July 6, 2015.
The observations were carried out under the queue observing mode using the fiber-fed ESPaDOnS 
echelle spectrograph equipped with an e2v 2048 x 4608 CCD (binned 1 x 1). The resolving power 
provided by this combination was about 80000 and the spectral range extended from 3700 to 
10500 \AA. The spectra were processed by the CFHT ESPaDOnS pipeline. The estimated S/N ratio 
at the continuum level depends upon the wavelength interval; typically this value is close to 
100.  Exposure times varied from 160, 1700 to 2400 seconds, from the brightest to faintest targets.

The processing of the spectra (continuum level determination, equivalent width measurements etc.) was 
carried out by using the DECH20 software package \citep{Gal92}. The list of spectral lines is the same as 
used in our previous studies (see \citealt{Kov99}). 

After the preliminary spectra processing, we discovered that three of our program stars (ASAS 175621-2738.0,    
ASAS 183406-1519.1 and ASAS 193435-1921.7) showed a very strong emission feature in their H$\alpha$ absorption profiles.  
This characteristic is considered to be a sure sign of W Vir variable type stars (see, e.g. \citealt{Sch04}). Thus these 
three stars were discarded from this study; they will be analysed in another paper.  The H$\alpha$ profile for ASAS 181024--2049.6 
is free of this type of emission and the light curve of this star in V band is typical of classic $\delta$ Cep type stars (868 data points 
with a good phase coverage, see ASAS database).  Additional support for this classification is ASAS 181024-2049.6's small Galactic 
latitude  ($b \approx -0.8^{\circ}$).

\section{Stellar atmosphere parameters}

The effective temperature (T$_{\rm eff}$) of ASAS 181024--2049.6 was estimated by calibrating 
the ratios of the central depths of the lines with different potentials at the lower levels (see \citealt{Kov07}).
This method gives an error in the temperature of about $\pm 150$ K (error of the mean is 20 K with 50 calibrating
relations used). The surface gravity was computed using the iron ionization balance. It produces uncertainty in the gravity
measurement of about $\pm 0.3$ dex. The microturbulent velocity  was found by avoiding any dependence between the iron 
abundance  as produced by individual Fe~II lines and their equivalent widths. Error in the V$_{\rm t}$ value is about $\pm 0.5 $km~s$^{-1}$. The resulting atmosphere parameters for ASAS 181024--2049.6 are listed in Table \ref{Par}.  

It should be noted that ASAS lists period of this star P= 24.43288 d and epoch of Max = 2029.5 that leads to estimated phase of 
0.026 at the date of observation, but this is in conflict with our spectroscopic estimate of T$_{\rm eff}$ = 5100 K, which is apparently 
too low for the phase near the maximum light. The reason for this discrepancy could be the wrong light elements for this star in ASAS database, or pulsation period change. No information about the possible period change for this stars is available in the literature.

\begin{table*}
\small
\caption{Some characteristics of our program star.}
\label{Par}
\begin{tabular}{cccccccc}
\hline
 Star           &  JD   & $<$V$>$ (ASAS) & T$_{\rm eff}$ [K] & $\log~g$&  V$_{\rm t} [$km~s$^{-1}]  $&  E(B--V) &  R$_{\rm G}$ [kpc] \\
\hline                                           
ASAS 181024--2049.6&  2457209.9111  & 13.525  & 5113 & 1.4 & 3.1 &  1.41 & 2.53 \\
\hline
\end{tabular}

Remark: the reddening value was found using a special method (see Sect. 5).

\end{table*}

\section{Spectroscopic analysis}

All abundances listed in Table \ref{Ab1}  (with the exception of oxygen, sulfur and barium) 
were derived in the LTE approximation in order to be consistent with our previous abundance 
determinations in Cepheids. For this we used the ATLAS9 code \citep{Kur92} to generate 
the appropriate atmosphere models, and Kurucz's WITH9 code to analyze the equivalent widths.
For comparison we used the solar chemical composition from \citet{Grevet96}.

In \citet{Maret15} we derived the NLTE abundance distributions for O, S and Ba for all our program
Cepheids (new stars and those previously investigated by \citealt{luc13}, \citealt{AND13}, 
\citealt{Kor14},  \citealt{AND14}).  All the details concerning the calculation method of the NLTE 
calculations and the atomic models as applied for oxygen, sulfur and barium can be found in those papers.

\begin{table}
\caption[]{Elemental abundances in ASAS 181024--2049.6}
\label{Ab1}
\begin{tabular}{lrrrr}                             
\hline
  Ion & [A/H] & $\sigma$ & NoL & (A/H) \\
\hline 
 6.00 &    0.02 & 0.02 &    2  &  8.57 \\
 7.00 &    0.66 & 0.03 &    2  &  8.63 \\
 8.00*&    0.46 & 0.12 &    5  &  9.17 \\
11.00 &    0.93 & 0.00 &    1  &  7.26 \\ 
12.00 &    0.49 & 0.11 &    3  &  8.07 \\
13.00 &    0.60 & 0.04 &    4  &  7.07 \\
14.00 &    0.46 & 0.07 &   29  &  8.01 \\
14.01 &    0.45 & 0.08 &    2  &  8.00 \\
16.00*&    0.44 & 0.15 &    8  &  7.60 \\
20.00 &    0.34 & 0.16 &    4  &  6.70 \\
21.00 &    0.35 & 0.11 &    4  &  3.52 \\
21.01 &    0.39 & 0.09 &    5  &  3.56 \\
22.00 &    0.39 & 0.10 &   52  &  5.41 \\
22.01 &    0.47 & 0.11 &    2  &  5.49 \\
23.00 &    0.46 & 0.07 &   28  &  4.46 \\
23.01 &    0.37 & 0.07 &    4  &  4.37 \\
24.00 &    0.41 & 0.09 &   28  &  6.08 \\
24.01 &    0.16 & 0.13 &    6  &  5.83 \\
25.00 &    0.54 & 0.15 &    8  &  5.93 \\
26.00 &    0.44 & 0.07 &  202  & 7.94  \\
26.01 &    0.44 & 0.08 &   29  &  7.94 \\
27.00 &    0.43 & 0.11 &   32  &  5.35 \\
28.00 &    0.35 & 0.04 &   26  &  6.60 \\ 
29.00 &    0.57 & 0.00 &    1  &  4.78 \\
37.00 &    0.38 & 0.00 &    1  &  2.98 \\
38.00 &    0.54 & 0.00 &    1  &  3.44 \\
39.01 &    0.30 & 0.24 &    6  &  2.54 \\
40.00 &    0.31 & 0.09 &    3  &  2.91 \\
40.01 &    0.33 & 0.11 &    4  &  2.93 \\
56.01*&    0.70 & 0.15 &    3  &  2.87 \\
57.01 &    0.29 & 0.17 &    3  &  1.51 \\
58.01 &    0.16 & 0.06 &    7  &  1.71 \\
59.01 &    0.19 & 0.18 &    3  &  0.90 \\
60.01 &    0.15 & 0.09 &    6  &  1.65 \\
63.01 &    0.53 & 0.00 &    1  &  1.04 \\
64.01 &    0.33 & 0.00 &    1  &  1.45 \\
\hline\\
\end{tabular}                       
                                                             	
Remarks: \\
     $\ast$ -- NLTE abundances,\\
     $[$A/H$]$ -- relative-to-solar abundance of element, \\ 
     NoL -- number of lines, \\
     (A/H) -- absolute abundance of element with respect to A(H)=12.00
     .
\end{table}

\section{Results and discussion}

\begin{figure}
\includegraphics[width=8.5cm]{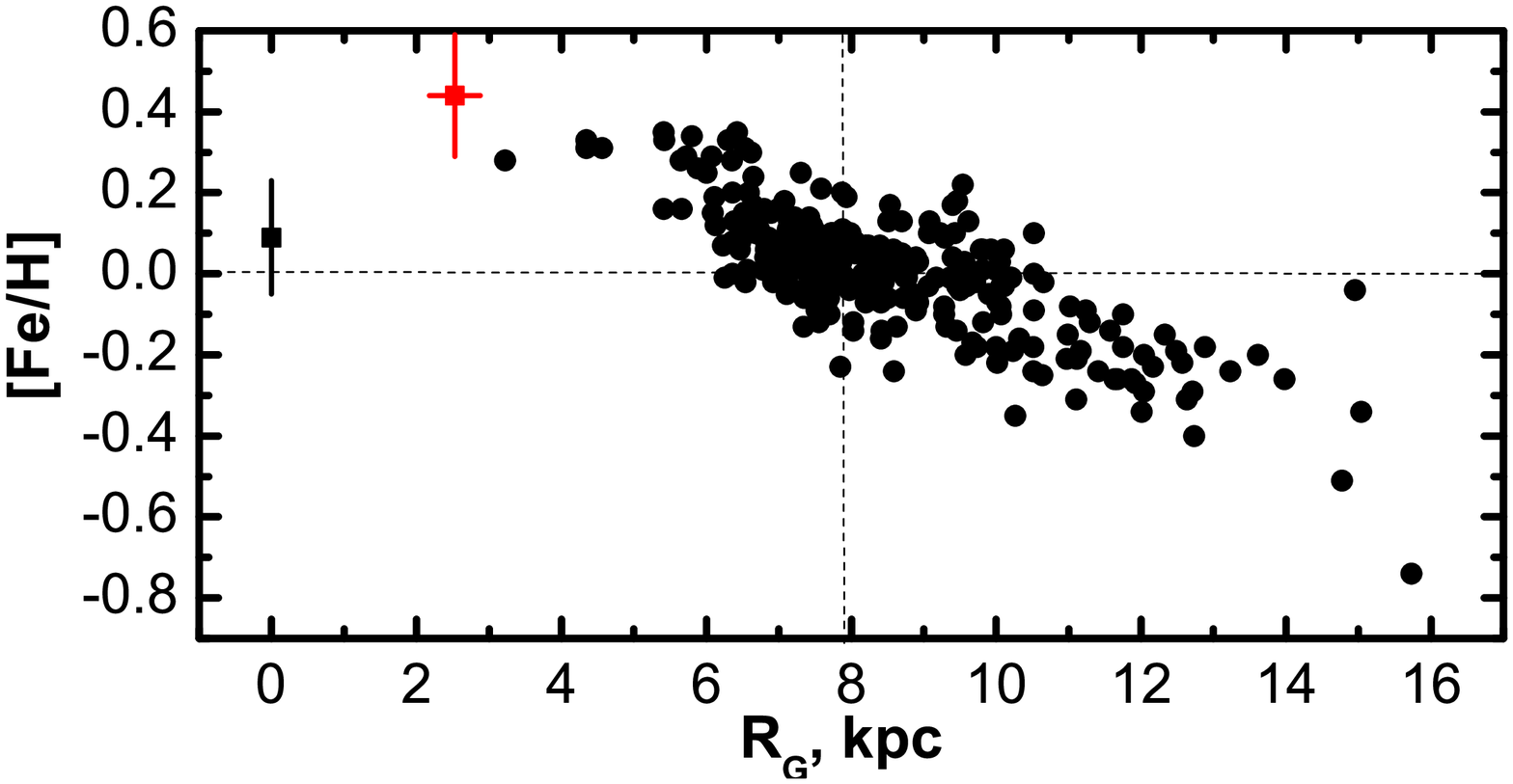}
\includegraphics[width=8.5cm]{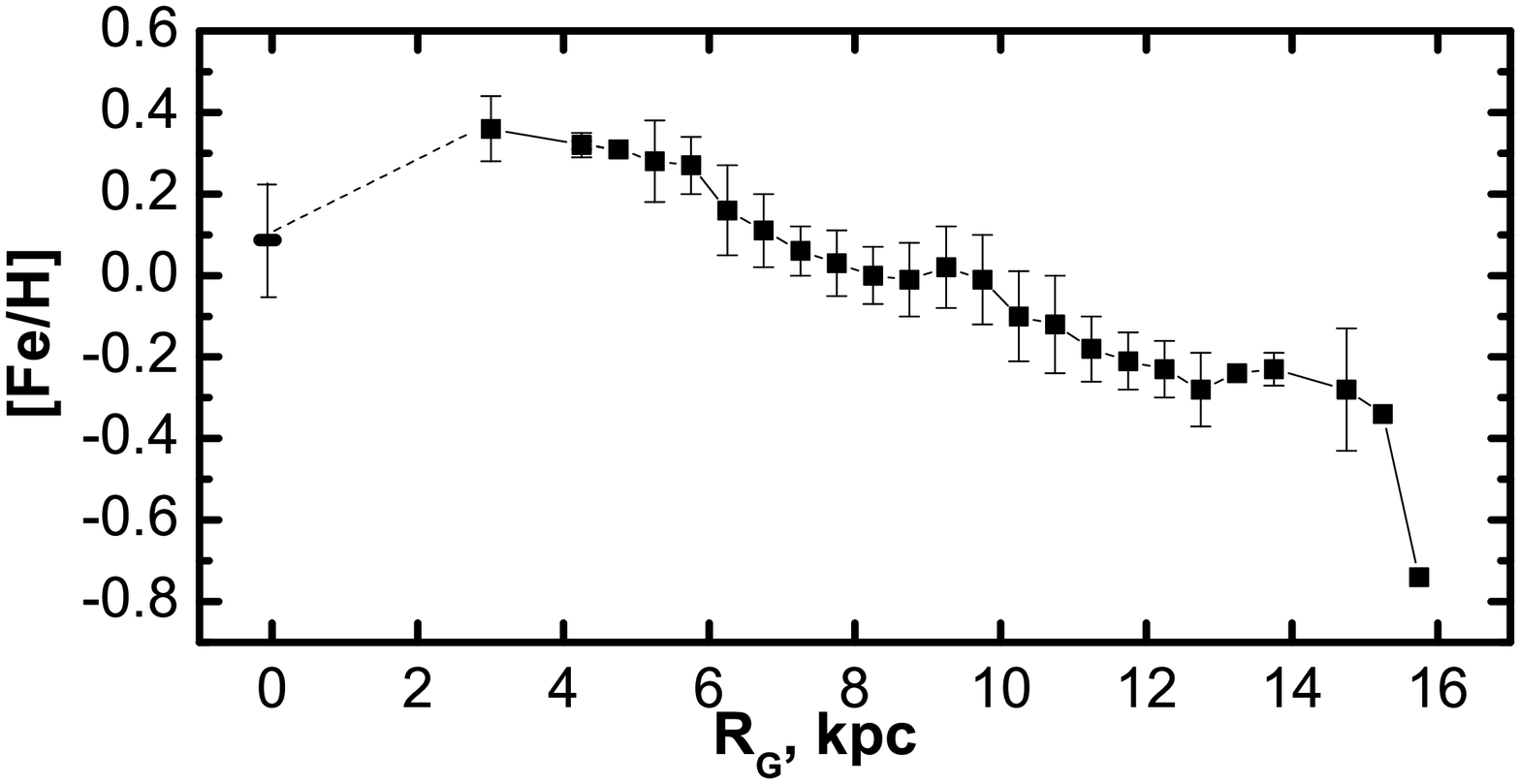}
\caption[]{$Top$: LTE iron abundance distribution in Galactic disc. Black 
circles - our previous determinations published in a series of papers 
mentioned in the Introduction. For ASAS 181024-2049 (red square) we 
show the typical iron abundance uncertainty for Cepheids ($2\sigma$).\\
$Bottom$: The same as in the upper plot but for binned data for the radial extent of the 
disc containing the bulk of the Cepheids. Our program star and the black circle at the center of the Galaxy
were simply connected with a dashed line.}
\label {FeRg}
\end{figure}

In Fig. \ref{FeRg} we show the iron abundance distribution of Galactic Cepheids from  all of our previous studies 
together with the Galactic center abundances as described in Martin et al. (2015). In order to include ASAS 181024-2049.6 
in this plot, we first need its heliocentric distance. Unfortunately,  the E(B--V) value for this star is currently unknown. 
To overcome this problem we applied the method of the colour excess determination based on the use of 
the E(B--V) -- EW(DIB) calibrating relation from \citet{Friedetal11}. Here EW(DIB) is the equivalent width of the 6613 \AA~ 
diffuse interstellar band (DIB), which is also seen in Cepheid spectra. Using this relation, we find that the E(B--V) 
value for ASAS 181024-2049.6 is equal to 1.41. Combining the approximate relation A$_{\rm v}$ = 
E(B--V)/3.2, the ''absolute magnitude--pulsational period" relation of Gieren et al. (1998), and the mean V 
magnitude from Table \ref{Par}, we find a heliocentric distance $d= 5641$ pc, and Galactocentric distance R$_{\rm G}= 
2.53$ kpc. 

With this Galactocentric distance, our program star falls within a very critical, under-sampled region of the 
inner disc of the Milky Way. Using that star and our Cepheid spectral analysis, we can better assess the distribution of abundances 
within the inner 3 kpc, as first explored by \citet{Maret15}. As one can infer from the plot, at Galactocentric distances less 
than 5--6 kpc, the iron abundance distribution seems to form a plateau-like structure. From our 
restricted data one can conclude that the maximal iron abundance in the thin disc perhaps reaches about +0.4 dex. 
New data could change this value, but the global behavior of the iron distribution in the disc is unlikely to be 
affected very much.

\begin{figure}
\includegraphics[width=8.5cm]{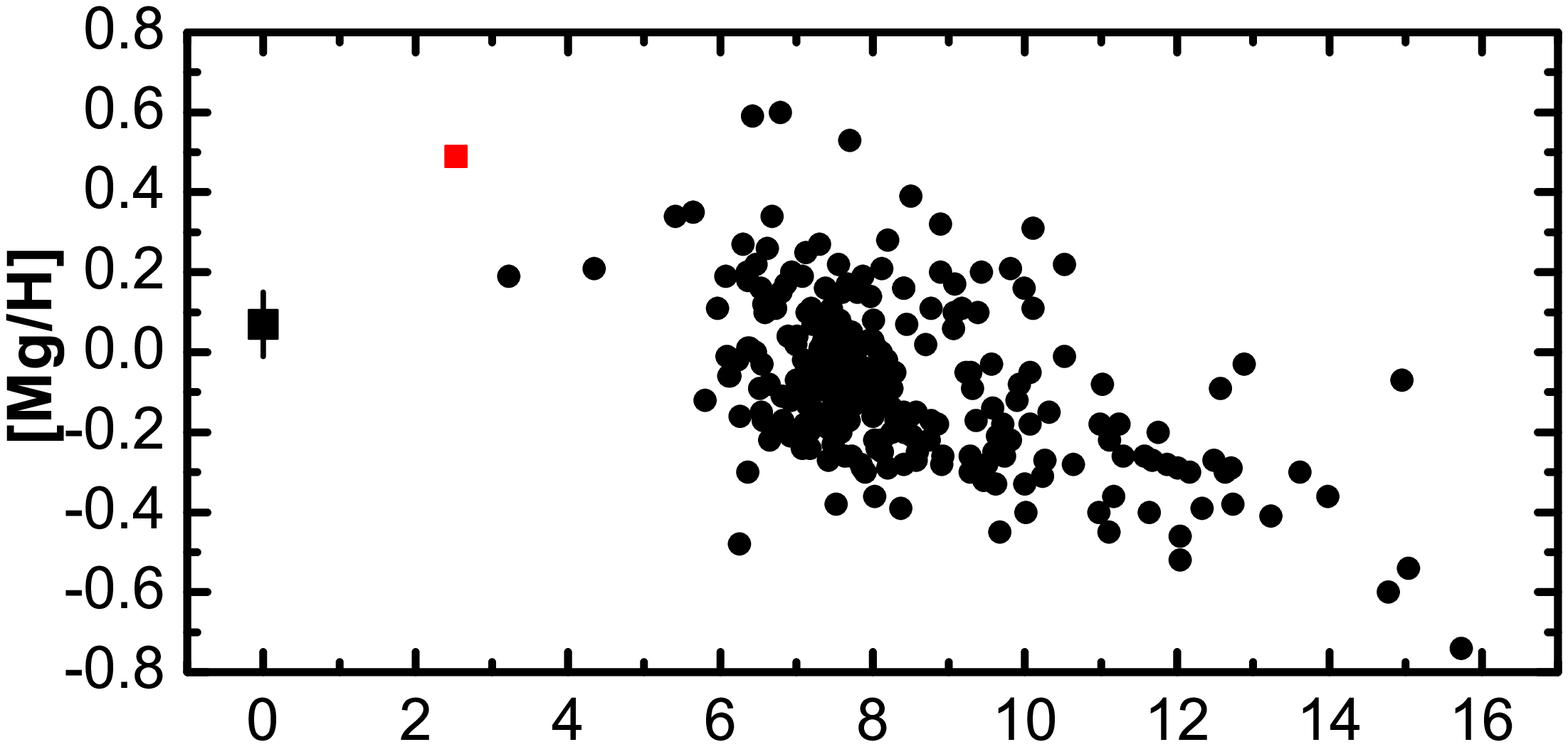}
\includegraphics[width=8.5cm]{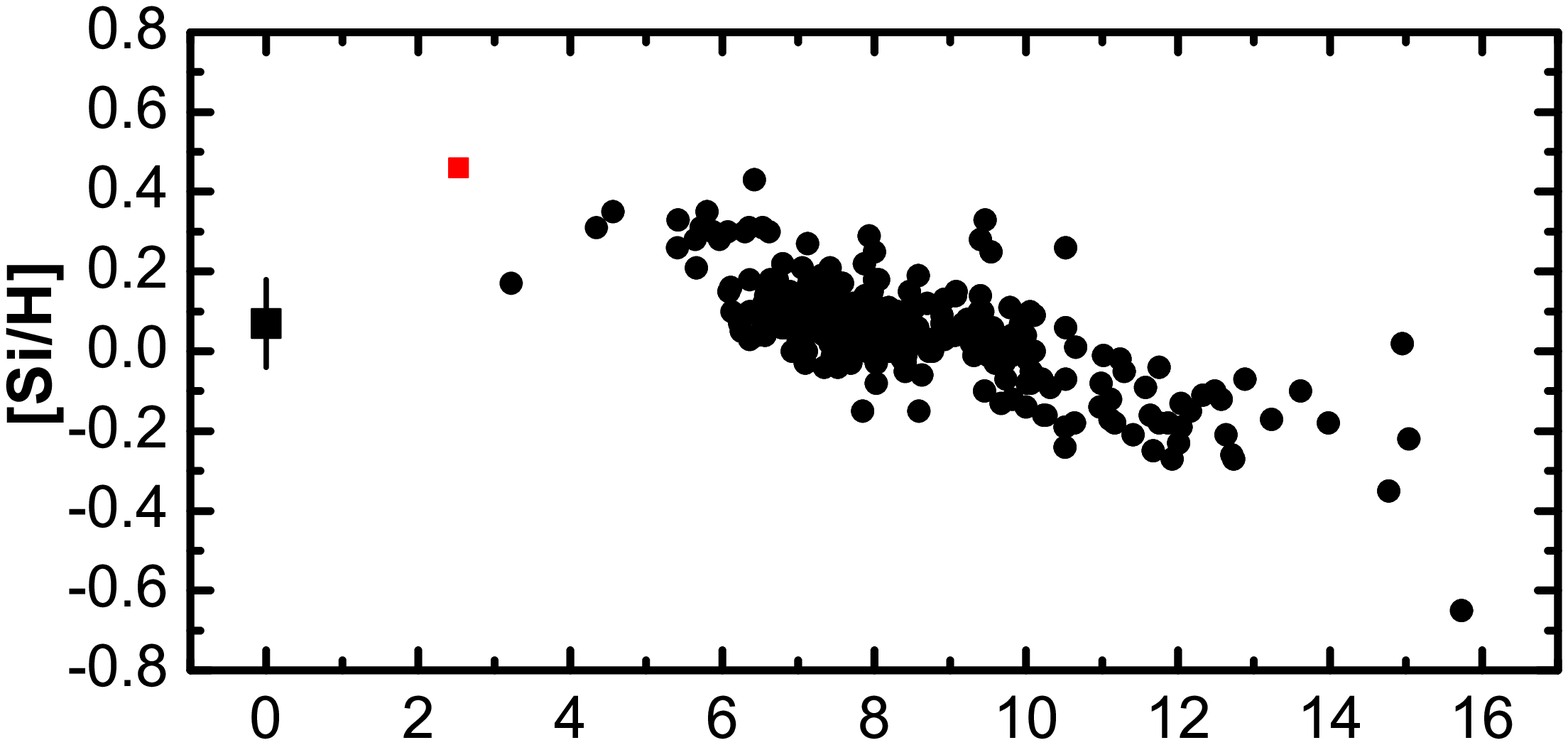}
\includegraphics[width=8.5cm]{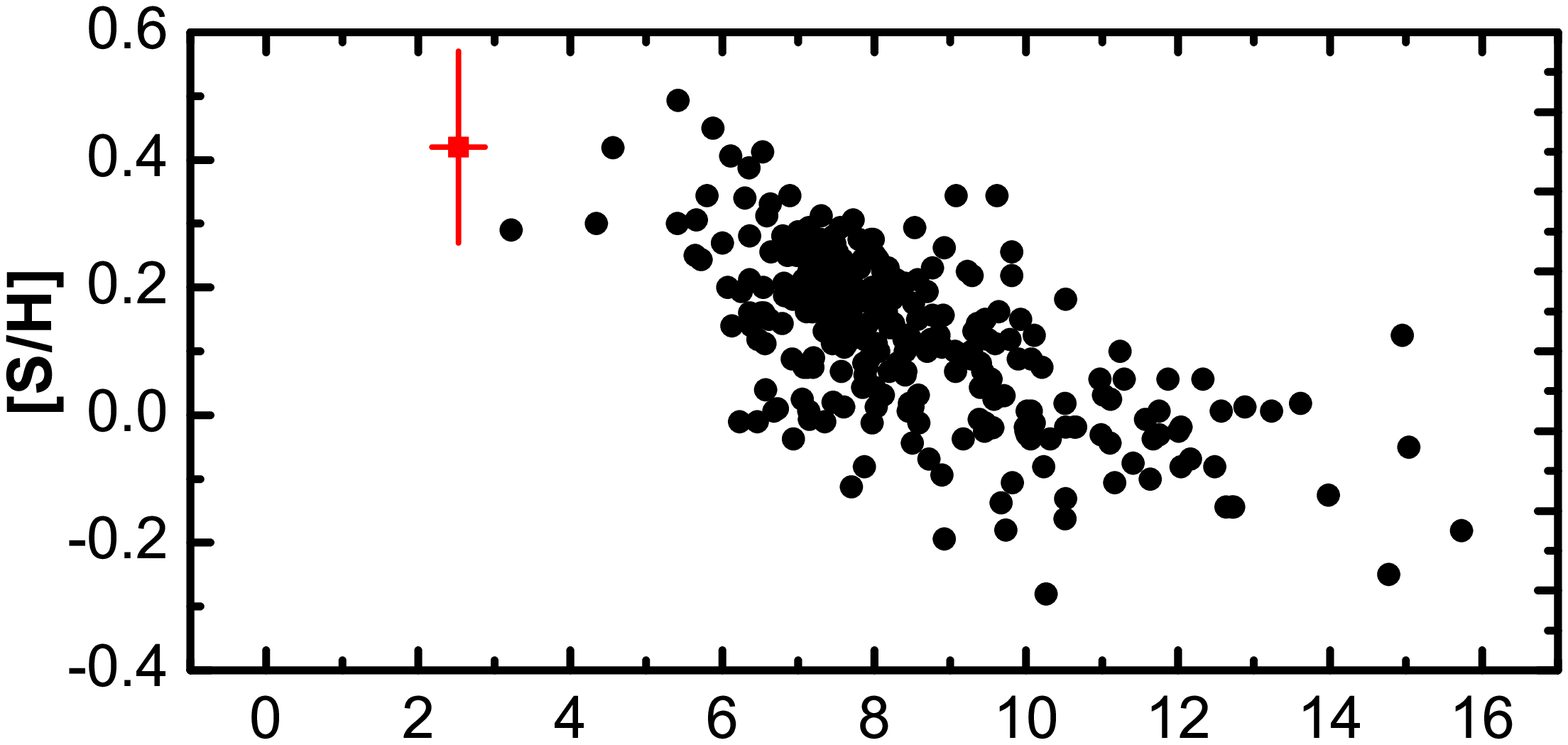}
\includegraphics[width=8.5cm]{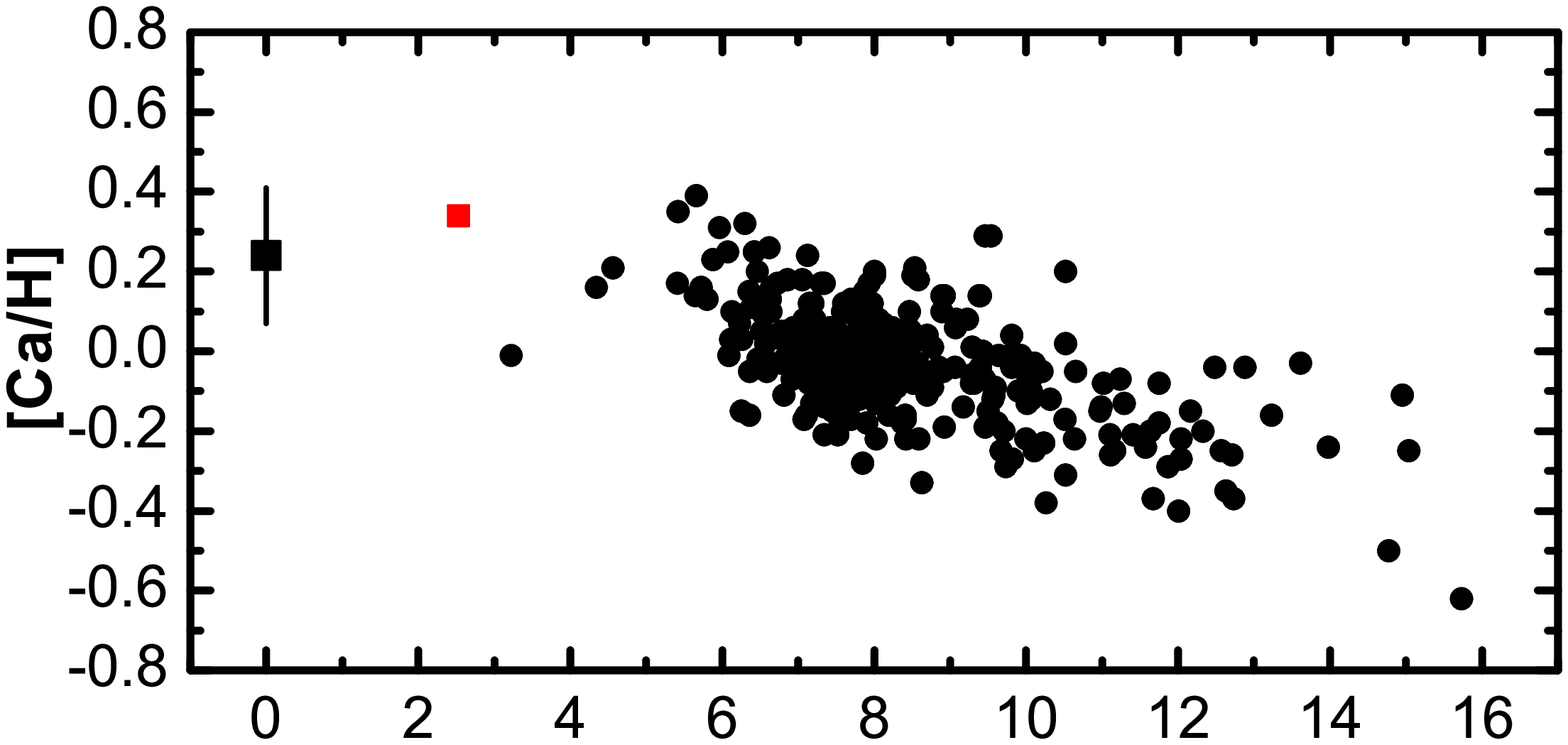}
\includegraphics[width=8.5cm]{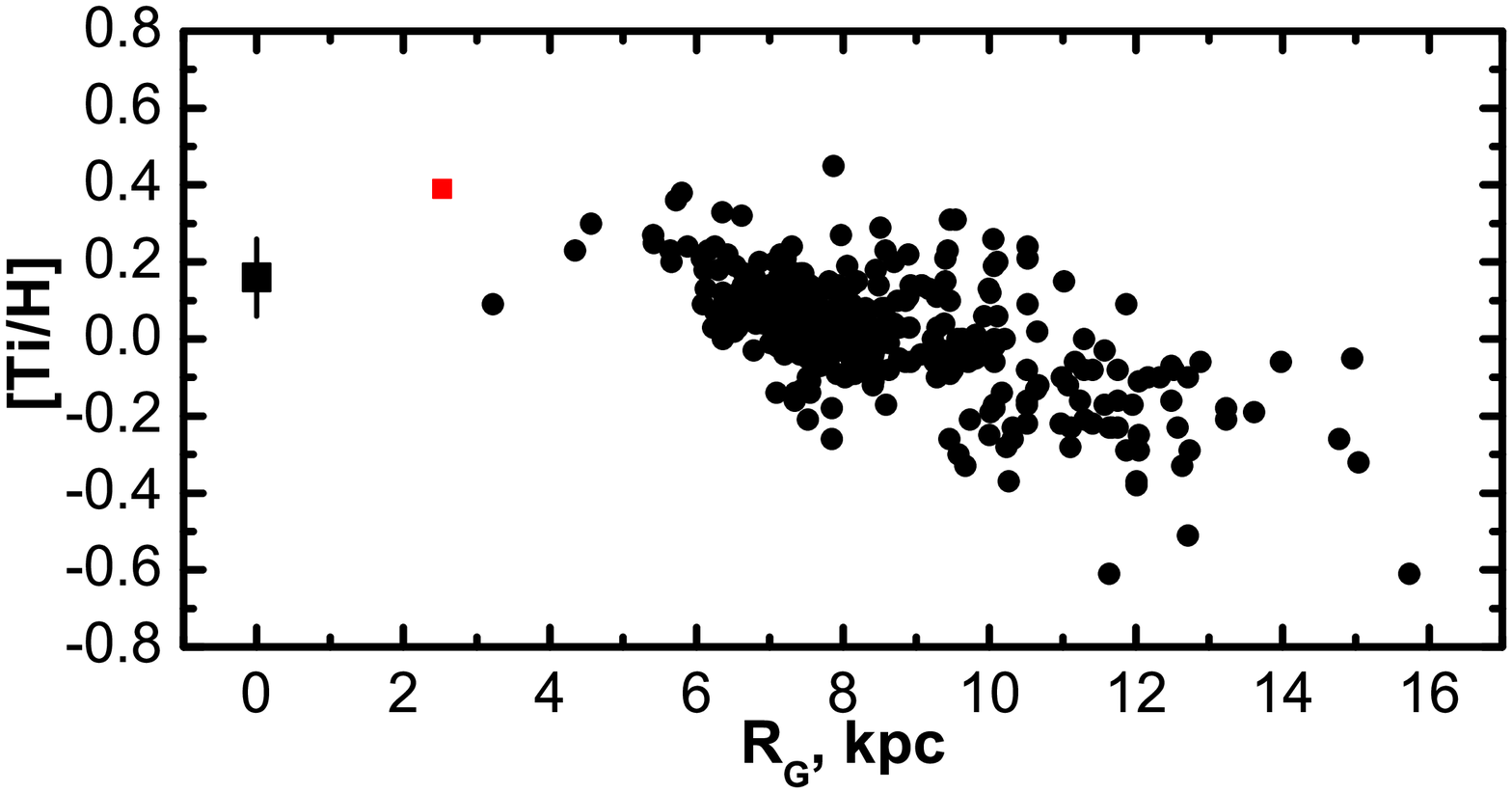}
\caption[]{The same as for Fig.1 but for magnesium, silicon, sulfur, 
calcium and titanium.}
\label {alRg}
\end{figure}

\begin{figure}
\includegraphics[width=8.5cm]{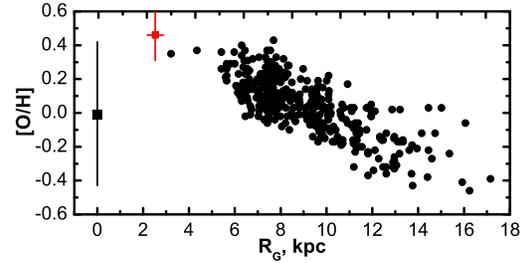}
\caption[]{The same as for Fig.1 but for NLTE oxygen abundances.}
\label {ORg}
\end{figure}

The same conclusion can also be made for other elements such as oxygen, magnesium, silicon, 
sulfur, calcium and titanium (see Fig. \ref{alRg} and \ref{ORg}).

The fact that the metallicity flattens or decreases when moving from a radius of 
5 kpc to the inner region should not be a surprise, although it does not fit a 
simple extrapolation of the gradient seen in the the 5-9 kpc range. All the 
chemical evolution models (see eg. \citealt{LF85}, \citealt{ChMG97}, 
\citealt{Mat08}) state that for a given Galactocentric radius, the rate of 
enrichment of metallicity is proportional to the star formation rate, $\psi$(t). 
The reason is that the chemical enrichment is due to the material restituted by 
the stars to the Interstellar Medium (ISM) at the end of their lives. More 
massive stars have shorter lives and provide the most efficient contribution. 
If $\psi$ is large, the number of massive stars formed per unit time is large, 
keeping the mass function constant, and the amount of enriched matter 
restituted to the ISM is large.

In addition to the metallicity enrichment, $\psi$ controls the number of 
existing stars in the thin disc. The stellar density is the result of the 
integration of the star formation rate over the lifetime of the disc. Of course, 
the most massive stars disappear along this process, but the less massive ones, 
which are the majority, remain there. Consequently, in a first approximation, 
ignoring  the merging of dwarf galaxies for instance, the stellar density and 
metallicity remain proportional. We argue that the metallicity decreases 
towards the center for the same reason that the stellar density is smaller, 
that is because  $\psi$ is smaller in that region than at larger radii along 
the history of disc formation.

In the past, models of the stellar density along the disc assumed a simple 
exponential decrease of the stellar density along the disc, with scale-lengths 
of 2-4 kpc. Inversely, this would imply quite a large density in the central 
regions.  However, the large density of stars in the central region is due to 
another Galactic component, the bulge.  The component that interests us here is 
the thin (or young) disc, to which the Cepheids belong.  Several decades ago, 
mathematical expressions based on fitting brightness profiles of external 
galaxies were proposed for instance by \citet{Free70} and \citet{Korm077}. 
These expressions describe galactic discs in which the disc component decreases 
both  towards the center and towards the outer regions, presenting a maximum at 
an intermediate radius. A recent star count model by \citet{Pol13} of our Galaxy 
was extensively verified by comparisons with the 2MASS catalogue. In this model 
the thin disc is represented by  a modified exponential as suggested by Kormendy, 
with a decrease of density towards the center. 

Why has the star formation rate been smaller in the inner kpc of the Galaxy thin 
disc? One possible reason is the lower density of gas available. In the CO survey 
of the Galactic disc by \citet{Damet87}, the authors show the 
profile of column density as a function of the galactic longitude, which 
presents peaks at longitudes 30$^\circ$ and 330$^\circ$ approximately 
(see their Figure 6). 
These longitudes correspond to tangential directions of a circle of 4 kpc from 
the center (8$\cdot$$\sin$($\it l$)). Inside this circle there is a minimum at 
about  $\pm 15^\circ$. This is interpreted as a molecular ring of maximum gas 
density, which extends to longitude 320$^\circ$ (radius 5 kpc)(see 
\citealt{Damet87}, Figure 6). Although this is not strictly a ring, it is a clear indication that there is a maximum density of gas at about 4--5 kpc and then a lower density inside this region. The lower gas density in the inner region is probably connected to the presence of the bar, a structure that possibly produces a flow of gas towards the Galactic center.

\section*{Acknowledgements}
SMA and SAK acknowledge the SCOPES grant No. IZ73Z0--152485 for financial support. 
We acknowledge the discretionary time allocated by Director D. Simons and the queue observing team at CFHT. Authors are also thankful to the anonymous referee for his/her valuable comments.


\begin{thebibliography}{}

       \bibitem[{Andrievsky} {et al.}(2013)]{AND13}
       Andrievsky, S.M., L\'epine, J.R.D., Korotin, S.A., Luck, R.E., Kovtyukh, V.V., 
       \& Maciel, W.J., 2013, \mnras, 428, 3252

        \bibitem[{Andrievsky} {et al.}(2014)]{AND14}
        Andrievsky, S.M., Luck, R.E., \& Korotin, S.A., 2014, \mnras, 437, 2106

       \bibitem[{Chiappini, Matteucci \& Gratton}(1997)]{ChMG97}
       Chiappini, C.; Matteucci, F.; Gratton, R., 1997, \apj, 477, 765

       \bibitem[{Dame} {et al.}(1987)]{Damet87}
       Dame, T. M.; Ungerechts, H.; Cohen, R. S.; de Geus, E. J.; Grenier, I. A.; May, J.; 
       Murphy, D. C.; Nyman, L.-A.; Thaddeus, P., 1987, \apj, 322, 706

    \bibitem[{Freeman}(1970)]{Free70}
    Freeman K. C., 1970, \apj, 160, 811

     \bibitem[{Friedman}(2011)]{Friedetal11}
    Friedman, Scott D.; York, Donald G.; McCall, Benjamin J.; Dahlstrom, 
    Julie; Sonnentrucker, Paule; Welty, Daniel E.; Drosback, Meredith M.; 
    Hobbs, L. M.; Rachford, Brian L.; Snow, Theodore P., 2011, ApJ 727, 33	


    \bibitem[{Galazutdinov} (1992)]{Gal92}
    Galazutdinov, G. A., 1992, Preprint SAO RAS, No. 92

     \bibitem[{Gieren et al.} (1998)]{Gierenetal98}
     Gieren, W. P., Fouqu\'e, P., G\'omez, M. 1998, \apj, 496, 17

    \bibitem[{Grevesse et al.}(1996)]{Grevet96}
    Grevesse N., Noels A., Sauval J., 1996, ASP Conf. Ser. 99, 117


     \bibitem[{Kashuba} (2016)]{Kas16}
     Kashuba, S.V., Andrievsky, S.M., Chekhonadskikh, F.A., Luck, R.E., Kovtyukh, V.V.,
     Korotin, S.A., Krelowski J., 2016, in preparation

     \bibitem[{Kormendy}(1977)]{Korm077}
     Kormendy J., 1977, \apj, 217, 406

     \bibitem[{Korotin} {et al.}(2014)]{Kor14}
      Korotin, S.A., Andrievsky, S.M., Luck, R.E., et al., 2014, \mnras, 444, 3301

     \bibitem[{Kovtyukh} (2007)]{Kov07}
     Kovtyukh, V.V., 2007, \mnras, 378, 617

    \bibitem[{Kovtyukh} \& {Andrievsky}(1999)]{Kov99}
     Kovtyukh, V.V., \& Andrievsky, S.M., 1999, \aap, 351, 597

     \bibitem[{Kurucz} (1992)]{Kur92}
     Kurucz, R. L. 1992, in The Stellar Populations of Galaxies, ed. B. Barbuy, 
      \& A. Renzini, IAU Symp. 149, 225

      \bibitem[{Lacey \& Fall}(1985)]{LF85}
      Lacey, C.G., Fall, S.M., 1985, ApJ 290, 154

       \bibitem[{Luck} {et al.}(2013)]{luc13}  
       Luck, R.E., Andrievsky, S.M., Korotin, S.N., \& Kovtyukh, V.V., 
       2013, \aj, 146, 18

       \bibitem[{Martin} {et al.}(2015)]{Maret15}  
       Martin R.P., Andrievsky S.M., Kovtyukh V.V., Korotin S.A., Yegorova I.A., Saviane Ivo, 
       2015, MNRAS 449, 4071

       \bibitem[{Matteucci}(2008)]{Mat08}
        Matteucci, F., 2008, IAUS 250, 391
        
        \bibitem[{Polido}{et al.}(2013)]{Pol13}
        Polido, P., Jablonski, F., L\'epine, J. R. D., 2013, \apj, 778, 32
        
        \bibitem[{Schmidt}{et al.}(2004)]{Sch04}
         Schmidt E.G., Johnston D., Lee K.M., Langan S., Newman P.R., Snedden S.A., 2004, 
        ApJ 128, 2988      

\end{thebibliography}
\end{document}